\documentclass[10pt]{article}
\usepackage[T1]{fontenc}
\usepackage{graphicx}
\usepackage{textcomp}
\usepackage{epstopdf}
\usepackage{hyperref}
\usepackage{graphicx}
\usepackage{amsmath}
\usepackage{amssymb}
\usepackage{float}
\usepackage{lipsum}

\newcommand{\h}{\hspace*{4ex}}

\newcommand{\cent}{\centerline}
\newcommand{\vs}{\vspace*}

\begin{document}

\baselineskip 0.5cm

\begin{center}

{\large {\bf Possibilities to generation of optical non-diffracting beams by holographic metasurfaces using surface impedance } }
%$^{\: (\dag)}$ } \footnotetext{$^{\: (\dag)}$ contact: marcos.gesualdi@ufabc.edu.br}

%%The authors acknowledge partial support from FAPESP (UNDER GRANTS 09/11429-2 and 11/51200-4); %%from CNPq (UNDER
%%GRANTS 307962/2010-5 and 309911/2011-7).

\end{center}

\vs{0.2 cm}

\cent{Santiago R. C. Fernandez$^{\: 1,2}$, and Marcos R. R. Gesualdi$^{\: 1}$}

\vs{0.2 cm}

\centerline{{\em $^{\: 1}$ Universidade Federal do ABC, Santo Andr\'e, SP, Brazil.}}
\centerline{{\em $^{\: 2}$ Universidad Nacional de Ingenieria, Lima, Peru.}}

\vs{0.5 cm}

{\bf Abstract  \ --} \ In this work, we present the computational simulations of holographic metasurfaces to realization of the optical non-diffracting beams. The metasurfaces are designed by the holographic technique and the computer-generated holograms (CGHs) of optical non-diffracting beams are generated computationally. These holographic metasurfaces (HMS) are obtained by modeling a periodic lattice of metallic patches on dielectric substrates with sub-wavelength dimensions, where each one of those unit cells change the phase of the incoming wave. We use the surface impedance (Z) to control the phase of the electromagnetic wave through the metasurface in each unit cell. The sub-wavelength dimensions guarantees that the effective medium theory is fulfilled. The results is according to the predicted by non-diffracting beams theory. These results are important given the possibilities of applications in optical tweezers, optics communications, optical metrology, 3D imaging, and others in optics and photonics.

%\h PACS nos.: 42.25.Bs; 42.25.Fx; 41.20.Jb; 46.40.Cd; 41.85.-p; 46.40.Cd; 42.20.Ht; 42.40.Jv, 42.30Lr.

%\author{XXX}

\vs{0.5 cm}

\h {\em 1. Introduction} --- The artificially structured materials such as photonic crystals and metamaterials have attracted great interest for their remarkable properties to control of light and electromagnetic waves \cite{ref:1,ref:2}. Metamaterials are artificial materials composed by a periodic array of sub-wavelength unit cells and they have been very explored in a wide range of applications due to fact of their properties depends on the geometry and materials of their unit cells \cite{ref:3,ref:4,ref:5,ref:6,ref:6a}. Some important applications of metamaterials are the obtaining of negative refractive index with resonant character \cite{ref:5,ref:7}, superlensing \cite{ref:7a,ref:7b}, the phenomena of negative refraction \cite{ref:8,ref:9,ref:10} and the possibility of cloaking light around certain physical spaces \cite{ref:11,ref:12}. The applications of three-dimensional metamateriais can be also applied to their two-dimensional versions: metafilms or metasurfaces \cite{ref:13,ref:14}. \\

Metasurfaces are interesting new devices to control or modify wavefront, phase or polarization state. The resonators introduce changes of phase in the interface due to the discontinuities on the surface. The result is a generalization of the laws of the reflection and refraction, being possible the control of a refracted wave by modulating the gradient of phase imposed by the resonators \cite{ref:15,ref:16, ref:17, ref:18,ref:19,ref:20}. \\

Holography was developed as a method for recording and reconstructing wavefronts such generating three-dimensional images \cite{ref:23,ref:24,ref:25}. Through recording process, the information of phase and amplitude of a wave scattered for the surface of an object (object wave) is stored in photosensitive materials (holographic recording medium) due interference with a reference wave \cite{ref:24,ref:25}. \\

On the other hand, non-diffracting waves are beams and pulses that keep their intensity spatial shape during propagation \cite{ref:26,ref:28,ref:29,ref:30,ref:31,ref:32,ref:33,ref:34,ref:35,ref:35a,ref:35b,ref:35c,ref:35d, ref:35e,ref:36}. Pure non-diffracting waves include Bessel beams, Mathieus beams, Parabolic beams, and others; as well as the superposition of these waves can produce very special diffraction-resistant beams, such as the Frozen Waves. The recent advances in generation of the non-diffracting beams by computational holography and spatial light modulators have possibilities many applications in many fields in photonics.\\

In this work, we use the surface impedance ($Z$) to control the phase of a wavefront through the metasurface by designing each unit cell in according with the hologram of this wavefront, such metasurface is called holographic metasurface (HMS). The holographic metasurfaces (HMS) to generation of non-diffracting waves in optical frequencies are computationally simulated. \\

\h {\em 2. Holographic principle and phase holograms} --- The holographic principle is an interferometric-diffractive technique and can be described as a recording and reconstruction process of the complete information of a wavefront, where the amplitude and phase information associated with the wavefront generated by the object is stored in an hologram\cite{ref:24,ref:25}. Techniques that involve computational recording and reconstruction processes are called computational holography, like computer-generated holograms (CGHs), which are related to the numerical recording of the hologram. CGHs can be used to reproduce the wave fronts of three-dimensional objects and optical waves with any prescribed amplitude and phase distribution. They are, therefore, extremely useful tool in applications such as the generation of optical beams and waves, optical processing, optical spatial filtering, 3D imaging, among others \cite{ref:23,ref:24,ref:25}. 
The amplitude and phase of a wavefront can be recorded using the transmittance function of a computer-generated phase hologram, given by \cite{ref:24}

\begin{equation}
 H\left(x,y \right)=\text{exp}\left[\psi\left(a,\phi \right)  \right]
\label{Phase_hologram}
\end{equation}

where $\psi\left(a,\phi \right)$ is the phase modulation of the computer-generated hologram and contains information on the amplitude and phase of the field. The equation \ref{Phase_hologram} can be expressed as a Fourier series in the domain of $\phi$, that is:

\begin{equation}
H\left(x,y \right)=\sum_{q=-\infty}^{\infty}H_{q}\left(x,y \right) 
\label{Fourier_Phase_hologram}
\end{equation}

and

\begin{equation}
H_{q}\left(x,y \right)=c_{q}^{a}\text{exp}\left(iq\phi\right) 
\label{Fourier_Phase_hologram_1}
\end{equation}

\begin{equation}
c_{q}^{a}=\dfrac{1}{2\pi}\int^{\infty}_{-\infty}\text{exp}\left[i\psi\left(a,\phi \right)  \right]\text{exp}\left(-iq \phi \right) d\phi 
\label{Coeficiente_Fourier_Phase_hologram}
\end{equation}

The first term of the equation \ref{Fourier_Phase_hologram} reconstructs the original field if the following identity is satisfied,

\begin{equation}
c_{1}^{a}=Aa
\label{Coeficiente_Fourier_Phase}
\end{equation}

$A$ is positive. \\

In this hologram, the phase modulation is given by,

\begin{equation}
\psi\left(a,\phi \right)=\phi + f(a)\text{sin}(\phi)
\label{Phase_modulation_type1}
\end{equation}

and the equation ~\ref{Phase_hologram} can be written as,

\begin{equation}
H\left(x,y \right)=\text{exp}\left[i \phi \right] \text{exp}\left[i f\left(a \right)\text{sin}\left(\phi \right) \right]
\label{Phase_Hologram_type1}
\end{equation}

Using the $Jacobi-Anger$ identity, 

\begin{equation}
 \text{exp}\left[if\left(a \right)\text{sin}\left(\phi \right)\right]= \sum_{m=-\infty}^{\infty}J_{m}\left[f\left(a\right)  \right]\text{exp}\left(im\phi\right)\,,
\label{Jacobi_Anger}
\end{equation}

where $ J_{m} $ correspond to the $ m $ Bessel functions. The equation \ref{Phase_Hologram_type1} can be written as,

\begin{equation}
 H\left(x,y \right)=\text{exp}\left[i \phi \right]  \sum_{m=-\infty}^{\infty}J_{m}\left[f\left(a\right) \right]exp\left(im\phi\right)
\label{Phase_Hologram1}
\end{equation}

and, from the equations \ref{Fourier_Phase_hologram} and  \ref{Fourier_Phase_hologram_1} the co-factors  $c_{q}^{a}$ are given by:  

\begin{equation}
 c_{q}^{a} = J_{q-1}\left[f\left(a\right)  \right] 
\label{Phase_Hologram2}
\end{equation}
of the relation \ref{Coeficiente_Fourier_Phase} we have that $J_{0}\left[f\left(a\right)\right]=a$ is satisfied for each value of $ a $, where $ A = 1 $ and is solved numerically to get the values $f(a)$. \\

The Fourier spectrum $ U(\xi, \eta) = \textit{F} \left \lbrace u(x, y) \right \rbrace $ of the encoded field $ u(x, y) $ is centered on the plane Fourier $ (\xi, \eta) = 0 $. Therefore the spectrum of the different terms $ H_{q} $ from Equation \ref{Fourier_Phase_hologram} are centered on the same plane $ (\xi, \eta) = 0 $. Thus, the encoded field cannot be recovered by spatial filtering. For spatial isolation of diffraction orders, the hologram is modified by adding a phase modulation $ 2 \pi (\xi_{0} x + \eta_{0} y) $. Therefore, the modified transmittance function is

\begin{equation}
 H\left(x,y \right)=\sum_{q=-\infty}^{\infty}H_{q}\left(x,y \right)\text{exp}\left[2\pi i \left(u_{0}x+ v_{0}y \right)  \right]
\label{Phase_Hologram_modificado}
\end{equation}

The Fourier spectrum of this modified CGH is given by:

\begin{equation}
 h\left(x,y \right)=\sum_{q=-\infty}^{\infty}h_{q}\left(u-qu_{0},v-qv_{0} \right)
\label{Phase_Hologram_modificado_Fourier}
\end{equation}

where $ h_{q}(u,v) $ is the Fourier transform of $ H_{q}(x, y) $.

\h {\em 3. Holographic metasurface by surface impedance} ---

In this holographic metasurface, the interference happens between the surface wave $\psi_{\textrm{surf}}$, the incoming wave passing through the surface (reference wave), and the radiation wave $\psi_{\textrm{rad}}$, the transmitted wave from the surface (object wave). For reconstructing the radiation wave, we use the surface wave to excite the interference pattern: $(\psi_{\textrm{surf}}^{*}\psi_{\textrm{rad}})\psi_{\textrm{surf}}=\psi_{\textrm{rad}}|\psi_{\textrm{surf}}|^{2}$. Thus, to realize the radiation pattern, we need a distribution for a surface impedance on the metasurface, the theoretical equation for the surface impedance is given by the interference of $\psi_{\textrm{surf}}$ and $\psi_{\textrm{rad}}$ \cite{ref:22,ref:40} 
\begin{equation}\label{e0}
Z=i\left[X+M\operatorname{Re}(\psi_{\textrm{rad}}\psi_{\textrm{surf}}^{*})\right]
\end{equation}
where $X$ and $M$ are modulation values.

\begin{figure}[H]
\centering
\includegraphics[width=10cm]{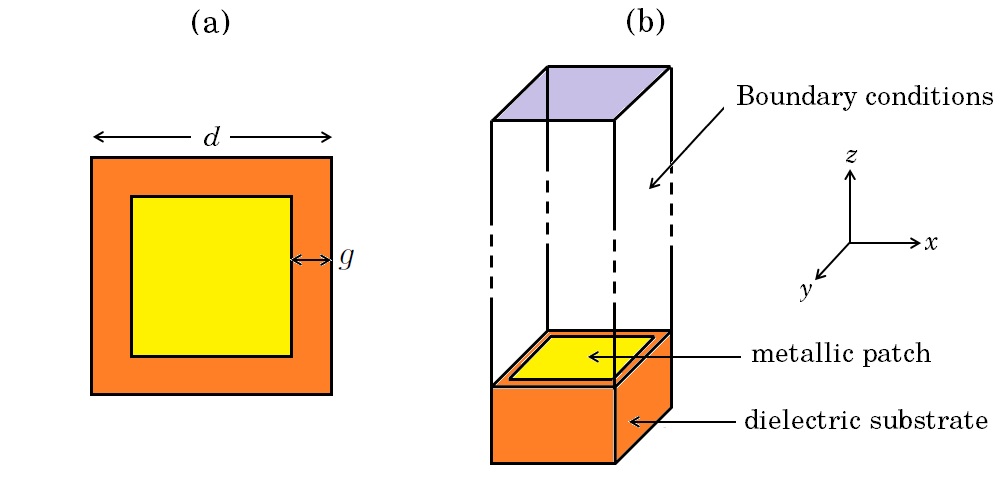}
\caption{\textbf{(a)} Unit cell of the metasurface. \textbf{(b)} Boundary conditions in the unit cell designed in CST.}
\label{f1}
\end{figure}

The impedance surface ($Z$) is defined as the ratio between the component of electric field parallel to a current along the surface and the current per unit length of surface: $\mathbf{E}_{t}=Z(\mathbf{x}_{t})\mathbf{J}$. For the case of a metasurface, we should average that equation over the unit cell, the result for the TM modes (transverse magnetic waves) is given by \cite{ref:22,ref:40}: 

\begin{equation}\label{e1}
Z=iZ_{0}\biggl(\frac{k_{z}}{k}\biggl)
\end{equation}

where $Z_{0}$ is the vacuum impedance, $k$ the wave vector and $k_{z}$ the decay constant, considering the surface wave as $Ae^{-i(\mathbf{k}_{t}\cdot\mathbf{x}_{t})-k_{z}z}\ e^{i\omega t}$. We can obtain another result for the surface impedance (\ref{e1}) by finding a value for the transverse wave vector, $k_{t}$. Using the software CST Microwave Studio, we can simulate a unit cell with lattice parameter $d$ and to find the phase ($\phi$) through such unit cell for the frequency $\omega$, according to $\phi=k_{t}d$: 

\begin{equation}\label{e2}
\left(\frac{k_{z}}{k}\right)^{2} = \left(\frac{k_{t}}{k}\right)^{2}-1 = \left(\frac{\phi/d}{\omega/c}\right)^{2}-1
\end{equation}

being $c$ the light speed at vacuum. Then, we can calculate the expression that relates the surface impedance and the phase through a unit cell: 

\begin{equation}\label{e3}
Z = Z_{0}\sqrt{1-\phi^{2}c^{2}/\omega^{2}d^{2}}
\end{equation}

In this way, we design the holographic metasurface (HMS) consisting on a set of unit cells, each one of them formed by a metallic patch on a dielectric substrate, both elements have square shape being the side of metallic patch always less than the side of substrate, thus, a gap ($g$) is formed in each unit cell (see Figure \ref{f1}) \cite{ref:40}. We can choose a determinate numbers of gaps for making the simulations, usually nine or ten values equally spaced are chosen, from a $g_{\textrm{min}}$ till $g_{\textrm{max}}$. For every value of gap, we design in CST Microwave Studio software the corresponding unit cell, applying the function Eingenmode solver to obtain the dispersion curve, i.e. the variation of frequency with the phase. \\

In this work, we want to obtain the holographic metasurfaces (HMS) of computer-generated holograms (CGH) of object waves previously calculated. Thus, the interference pattern in the expression \ref{e0} would be in the own CGH, i.e. the information of phase is given in each pixel of CGH. Every pixel of the computer-generated hologram has a gray level between 0 (black) and 255 (white), we associate each one of these values to a value of phase between 0 and $2\pi$ for obtaining a matrix of phase ($\Phi$) for the selected CGH. Thus, according to the expression \ref{e0}, we have \cite{ref:40}: $Z = i[X+M\Phi]$, where $X$ and $M$ would be adjustment values for making $Z$ to be inside the interval previously calculated for surface impedance $[Z_{\textrm{min}}, Z_{\textrm{max}}]$, making the calculations, we found: $X=Z_{\textrm{min}}$ and $M$ is the maximum integer value satisfying: $M\leq(Z_{\textrm{max}}-Z_{\textrm{min}})/2\pi$ \cite{ref:40}. \\

\h {\em 4. Non-diffracting beams} --- 

Non-diffracting waves are beams and pulses that keep their intensity spatial shape during propagation \cite{ref:26,ref:28,ref:29,ref:30,ref:31,ref:32,ref:33,ref:34,ref:35,ref:35a,ref:35b,ref:35c,ref:35d,ref:36}. \\

\textbf{Bessel beam}. The Helmholtz equation describes the propagation of waves taking into account the effects of diffraction and dispersion. Bessel beam is a solution of the Helmholtz equation with non-diffracting properties by choosing a coordinate system with a cylindrical circular cross-section $(\rho, \theta)$ \cite{ref:29}. In this coordinate system the transversal Helmholtz equation is given by,

\begin{equation}
\dfrac{1}{\rho}\dfrac{\partial}{\partial \rho}\left( \rho \dfrac{\partial \Psi_{\bot}}{\partial \rho}\right) + \dfrac{1}{\rho^2}\dfrac{\partial^2  \Psi_{\bot}}{\partial \theta^2} + k_{\bot}^2 \Psi_{\bot}=0.
\label{Helmholtz_trasnversal}
\end{equation}

Separating the solution as a product of two independent functions $\Psi_{\bot}\left(\rho,\theta \right) =\text{R}\left(\rho \right) \Theta \left(\theta \right)$, we get the system of independent differential equations

\begin{equation}
\rho \dfrac{d}{d\rho}\left(\rho \dfrac{df\left(\rho \right)}{d\rho} \right) + \left[\rho^2 k_{\bot}^2 -n^2 \right]\text{R}\left(\rho \right)=0\,,
\label{eqdiferencias1}
\end{equation}

\begin{equation}
\dfrac{d^2 \Theta \left( \theta \right)}{d\theta^2}-n^2 \Theta \left( \theta \right)=0\,,
\label{eqdiferencias2}
\end{equation}

where $n$ is a coupling constant. 

%The solution of the equation having an angular dependence is of the form $\text{e}^{in \theta}$. Since the medium is homogeneous and isotropic, the field is necessarily periodic with respect to $\theta$ and the value for the coupling constant $n$ is limited to integer value ​​$n=0, \pm{1}, \pm{2}, ... $. This equation, satisfied by the radial function $\text{R} \left(\rho \right)$, is the differential equation of the Bessel function and its particular solution is the Bessel function of the first type of order $n$, $J_{n} \left(k_{\bot} \rho \right)$, where $k_{\bot} = \sqrt{k_{0}^2-k_{z}^2} $ is the transverse wave number. Therefore, the complete solution to the Helmholtz equation for a wave propagating in the positive direction of the $z$ axis is of the form,

\begin{equation}
\Psi\left(\rho, \theta, z \right)=J_{n}\left(k_{\bot} \rho \right)\text{e}^{in \theta}\text{e}^{-ik_{z}z}
\label{solcompleta}
\end{equation}

\begin{figure}[H]
\centering
\includegraphics[width=8.1cm]{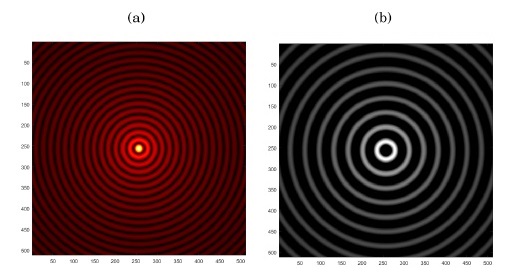}
\caption{(a) Amplitude, and (b) phase generation of the Bessel beam using MatLab algorithm.}
\label{f4a}
\end{figure}

From the equation (21) we see that the field strength is independent of the propagation distance $z$, ie $\vert \psi \left (\rho, \theta, z = 0 \right) \vert^2 = \vert \psi \left (\rho, \theta, z \right) \vert^2 = \vert J_{n} \left (k_{\bot} \rho \right) \vert^2$, hence the definition of non-diffracting. The intensity and phase profile can be seen in the Figure ~\ref{f4a}.

\textbf{Airy beam}. The solution for the Airy beam can be obtained through the paraxial diffraction equation in $(1+1)D$, which describes the dynamic propagation of the complex scalar amplitude $\psi$ of the electric field associated with planar beams \cite{ref:26},

\begin{equation}
i\dfrac{\partial}{\partial \xi}\psi\left(s,\xi \right)+\dfrac{1}{2}\dfrac{\partial^{2}}{\partial s^{2}}\psi\left(s,\xi \right)=0\,,
\label{paraxial}
\end{equation}

$s=x/x_{0}$ being a dimensionless transversal coordinate, $x_{0}$ an arbitrary scale factor, $\xi=z/kx_{0}^{2}$ a normalized distance, $z$ the coordinate along the propagation, $k=2 \pi n / \lambda_{0}$ the wave vector and $n$ the refractive index of the medium. The equation \ref{paraxial} admits a solution at $z=0$, given by 

\begin{equation}
\psi(s,0)=Ai(s)\,,
\label{Airy_1}
\end{equation}

where $ Ai $ is the Airy function and has the following integral representation 

\begin{equation}
Ai\left[s \right]=\dfrac{1}{2\pi}\int_{-\infty}^{\infty}\text{exp}\left[i\left(kx+\dfrac{k^{3}}{3} \right) \right]{dk}. 
\label{Airy_integral}
\end{equation}

The scalar field $ \psi \left (s,\xi \right) $ is obtained from the Huygens-Fresnel integral, which is highly equivalent with the equation \ref{paraxial}, and determines the field at a distance of $ z $ as a function of the field at $ z = 0 $,

\begin{equation}
\psi\left(s,\xi \right) =Ai\left(s-\dfrac{\xi^{2}}{4}\right)exp\left(i\dfrac{s \xi}{2} - i\dfrac{\xi^{3}}{12} \right)\,,   
\label{Airy_Propagation}
\end{equation}

this solution is known as the Airy beam, where the first term corresponds to the function $ Ai \left (s \right) $ shifted by $ \xi^{2} / 4 $ to the right, thus giving the parabolic trajectory to the beam. We can find the intensity by taking the square module of the field

\begin{equation}
I\left(s,\xi \right)=\vert \psi\left(s,\xi\right)\vert ^{2}=\left[  Ai\left(s-\dfrac{\xi^{2}}{4}\right)\right]^{2}.
\label{intensidadeAiry}
\end{equation}

\begin{figure}[H]
\centering
\includegraphics[width=8cm]{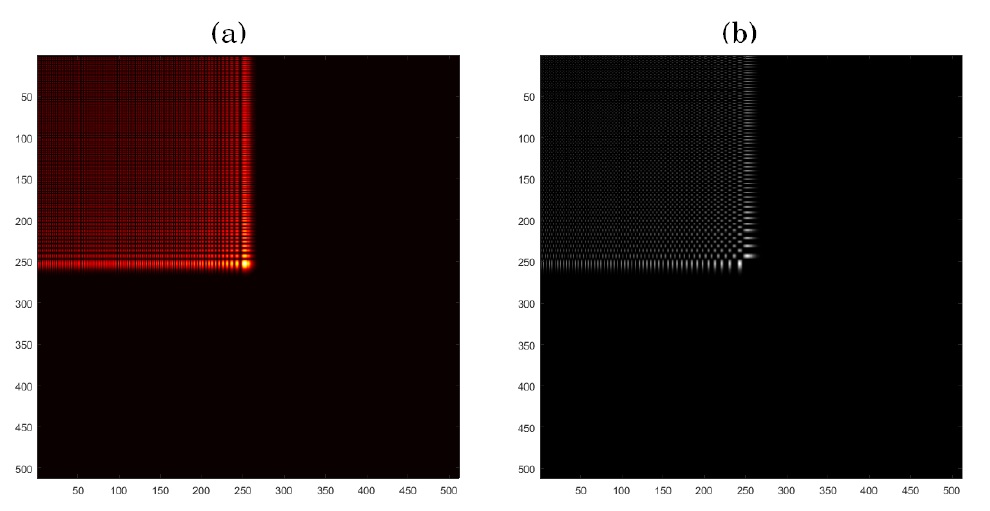}
\caption{(a) Amplitude, and (b) phase generation of the Airy beam using MatLab algorithm.}
\label{f5a}
\end{figure}

From the equation \ref{intensidadeAiry} we see that the cross-sectional intensity profile is due an Airy function, and remains unchanged during propagation while it undergoes constant acceleration. That is, the energy flow through a surface transverse to the direction of propagation is infinite, thus showing the non-diffracting property. \\

In the real case, where Airy beams propagate with finite energy, they were theoretically studied in the context of optics. For the construction of a beam that propagates with finite energy, a factor is introduced that exponentially truncates the field, equation \ref{Airy_1}, at $ z = 0 $ we will have 

\begin{equation}
\psi\left(s,0 \right)=Ai\left(s \right)\text{exp}\left(as \right)\text{exp}\left(i\nu s \right)\,,
\label{Airy_finity}
\end{equation} 

where the parameter $ a $ is a positive quantity that guarantees the convergence of the equation \ref{Airy_finity}, thus limiting the infinite energy of the beam, and the parameter $ \nu $ is associated with the initial launch angle $ \theta = \nu / kx_{0} $ of the parabolic trajectory. Under these initial conditions and from the Huygens-Fresnel diffraction integral, the field will evolve according to 

\begin{equation}
\begin{split}
\psi\left( s, \xi \right)&=Ai\left(s-\dfrac{\xi^{2}}{4}-\nu \xi + ia\xi\right)\text{exp}\left[ a \left(s - \dfrac{\xi^2}{2} - \nu \xi \right)\right] \\
	& \times  \text{exp}\left[ i \left( - \dfrac{\xi^3}{12} + \left(a^2 -\nu^2 + s \right)\dfrac{\xi}{2}  + \nu \xi -\nu \dfrac{\xi^2}{2} \right)\right].
\end{split}
\label{Airy_Balystic}
\end{equation}

Taking the intensity that describes the beam

\begin{equation}
I\left(s,\xi \right)=\vert \psi\left(s,\xi\right)\vert ^{2}=\left[ Ai\left(s-\dfrac{\xi^{2}}{4}-\nu \xi+ia\xi  \right)\right] ^{2}\exp\left(2as-a\xi^{2}-2\nu \xi \right)\,, 
\label{intensidadefinite}
\end{equation}

We see that Airy beam follows a ballistic trajectory in the $ s $ - $ \xi $ plane described by the parable $ s = \nu \xi + \xi^{2} / 4 $, and the intensity profile declines exponentially as result of modulation with the spatial exponential function in the initial plane $ \xi = 0 $. The term $ s_{0} = s- \left (\xi^{2} / 4 \right) - \nu \xi $, where $ s_{0} $ denotes the initial position of the peak at $ \xi = 0 $, defines the transverse acceleration of the peak intensity of the Airy beams and describes the parabolic path. \\

These results can be generalized in two dimensions 2D taking the scalar field that describes the beam as the product of two independent components, 

\begin{equation}
\psi\left(s_{x},s_{y},\xi_{x},\xi_{y}\right)=\psi_{x}\left(s_{x},\xi_{x}\right)\psi_{y}\left(s_{x},\xi_{y}\right)\,,
\label{Airy_2D}
\end{equation}

where each of the components $\psi_{x} \left (s_{x}, \xi_{x} \right)$ and $\psi_{y} \left (s_{x}, \xi_{y} \right)$ satisfies the equation \ref{paraxial}, $ s_{x} = x/x_{0} $, $ s_{y} = y/y_{0} $, $ \xi_{x} = z /kx_{0}^{2} $ e $ \xi_{y} = z / ky_{0}^{2} $. For a symmetric description of Airy beams the following conditions must be met: $ a_{x} = a_{y} = a $, and $ x_ {0} = y_{0} = w_{0} $, resulting in $ \xi_{x} = \xi_{y} = \xi = z / kw_{0}^{2} $. The intensity and phase profile can be seen in the figure ~\ref{f5a}. \\

\begin{figure}[H]
\centering
\includegraphics[width=8.3cm]{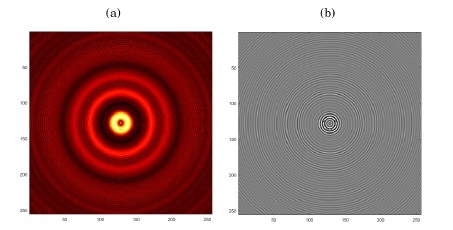}
\caption{(a) Amplitude, and (b) phase generation of the FW beam using MatLab algorithm.}
\label{f6a}
\end{figure}

\textbf{Frozen Wave (FW)}. The FW can be obtained by considering a superposition of $2N + 1$ beams order Bessel $ \nu $ with the same frequency $ \omega_{0} $ and different number of longitudinal wave $ k_{n} $ \cite{ref:30,ref:31,ref:32,ref:33,ref:34},

\begin{equation}
\Psi\left(\rho,\theta,z,t \right)=e^{-i\omega_{0}t} \sum_{n=-N}^{N} A_{n}J_{\nu}\left(k_{\rho n}\rho \right)e^{i \nu\theta} e^{-ik_{n}z} 
\label{FW}
\end{equation}

where $k$ is the total wave number, $k_{\rho n}$ is the transversal wave number and $k_{zn}$ is the longitudinal wave number for each value of $n$. The relation between them is: $k_{\rho n}^{2}=k^{2}-k_{zn}^{2}$, and $A_{n}$ are constant coefficients. 
Making the following choice for $ k_{zn} $, 

\begin{equation}
k_{zn}=Q+\frac{2\pi}{L}n
\label{beta}
\end{equation}

where $ Q>0 $ is the value to be chosen according to the experimental situation and obeys the following relation

\begin{equation}
0\leq Q+\frac{2\pi}{L}n \leq \frac{\omega}{c}
\label{beta}
\end{equation}

for $ -N \leq n \leq N $. This inequality determines the maximum value of $ n $, once $ Q $, $ L $ and $ \omega $ are chosen, so the equation \ref{FW} can be written as,

\begin{equation}
\Psi\left(\rho,\theta,z,t \right)=e^{-i\omega_{0}t}e^{-iQz} \sum_{n=-N}^{N} A_{n}J_{\nu}\left(k_{\rho n}\rho \right)e^{i\nu\theta} e^{-i\frac{2\pi}{L}nz}   
\label{FW4}
\end{equation}

with

\begin{equation}
A_{n}=\frac{1}{L}\int_{0}^{L}F(z)e^{-i\frac{2\pi}{L}nz}dz
\label{FW3}
\end{equation}

where $ \vert F(z) \vert^2 $ is the desired longitudinal intensity pattern in the range of $ 0 \leq z\leq L $. This intensity pattern can be concentrated on the propagation axis with $ (\rho = 0) $ when $ \nu = 0 $, or form a cylindrical surface when $ \nu > 0 $. The intensity and phase profile can be seen in the Figure ~\ref{f6a}. \\

In optics, the construction of the non-diffracting beam hologram is done numerically by a computer generated hologram (CGH) and its reconstruction is performed optically with its implementation in a spatial light modulator (LC-SLM). In this work, we will focus on computer-generated holograms of phase \cite{ref:25} of some types of non-diffracting waves. The non-diffracting waves are solutions to the (linear) wave equation which travel well confined or {\em localized\/}, in a single direction without to experiment effects of dispersion caused by diffraction. The types of non-diffracting waves studied are Airy beams \cite{ref:26,ref:35}, Bessel beams \cite{ref:28,ref:29} and Frozen waves (FW) \cite{ref:30,ref:31,ref:32,ref:33,ref:34,ref:35,ref:36}. 

\begin{figure}[H]
\centering
\includegraphics[width=14cm]{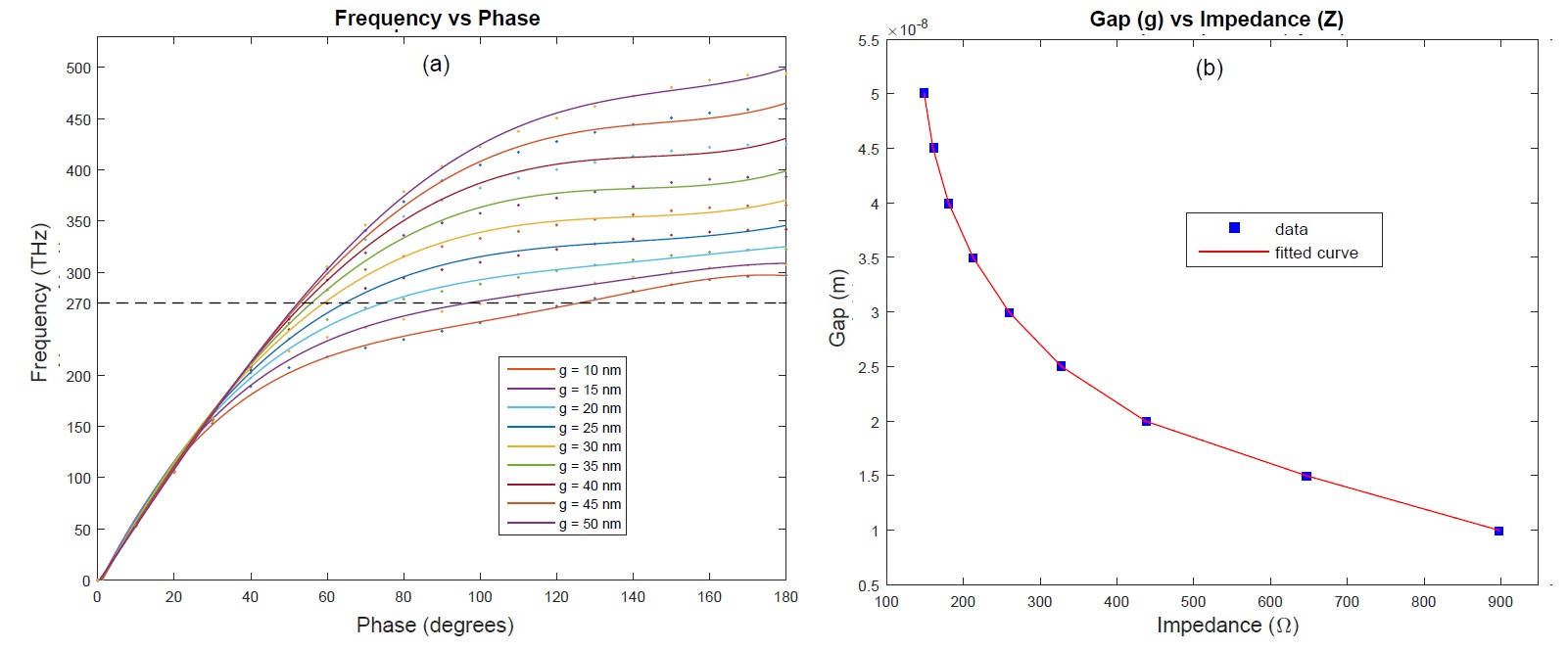}
\caption{(a) Variations of frequency with phase for each value of gap ($g$), the operating frequency was defined to 270THz ($\lambda=1.11 \mu m$) corresponding to region of near-IR. (b) Variation of values of gaps according to surface impedances and its corresponding adjustment curve.}
\label{f2}
\end{figure}

\h {\em 5. Results} --- We built three holographic metasurfaces working for different optical frequencies. The first of them, working at operating frequency of 270THz, the metal used was Aluminium (Al) on a substrate of Silicon (Si) with permittivity $\epsilon=11.9$ and thickness $t=40.5$nm. The lattice parameter was set at $d=150$nm, and gaps were setted from $g=10$nm till $g=50$nm with an interval of 5nm. The frequency was defined from 0 till 500THz. In the four sides of the unit cell (see Fig. \ref{f1}) were set periodic boundary conditions and PEC for the propagation direction. The graph of the dispersion curves for each value of gap and variation of gaps with surface impedance are shown in the Figure \ref{f2}. After making the adjustment of the curve of $g$ vs $Z$, we obtained the interval for surface impedance: $[Z_{\textrm{min}}, Z_{\textrm{max}}] = [149\Omega, 897.6\Omega]$, and we also found the following modulate values: $X=149\ \Omega$, $M=119$.

\begin{figure}[H]
\centering
\includegraphics[width=12cm]{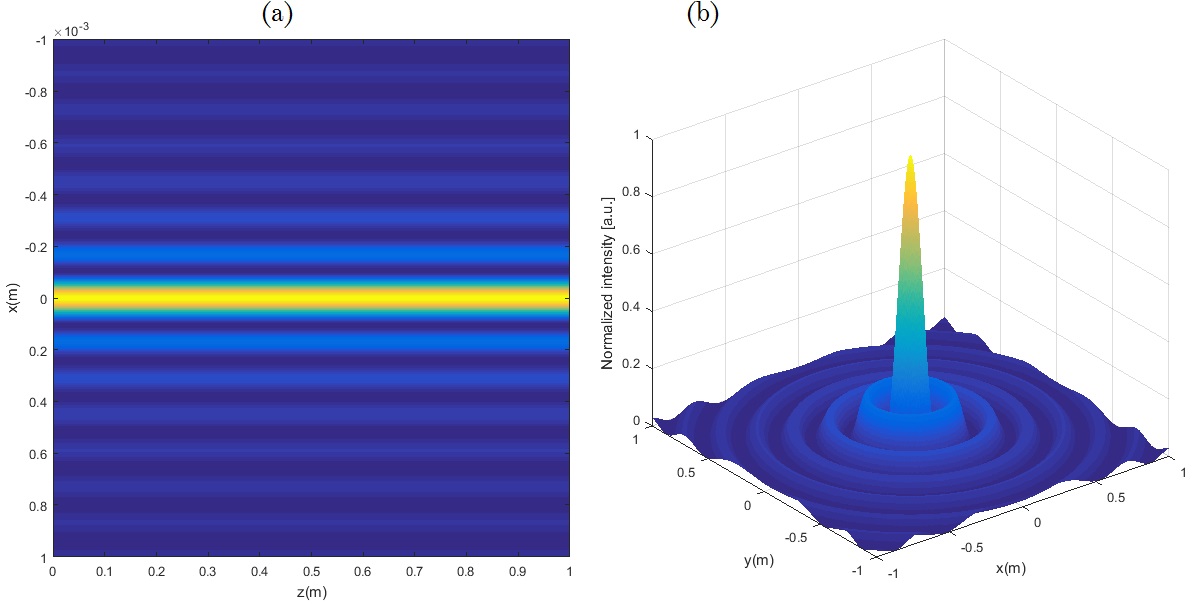}
\caption{(a) Orthogonal projection, and (b) transverse profile of the Bessel beam.}
\label{f4b}
\end{figure}

\begin{figure}[H]
\centering
\includegraphics[width=7cm]{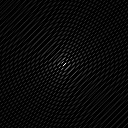}
\caption{Image of the computer-generated hologram of a Bessel beam with resolution of 128x128 pixels and at wavelength $\lambda=12.33$ mm.}
\label{f4}
\end{figure}

\begin{figure}[H]
\centering
\includegraphics[width=12cm]{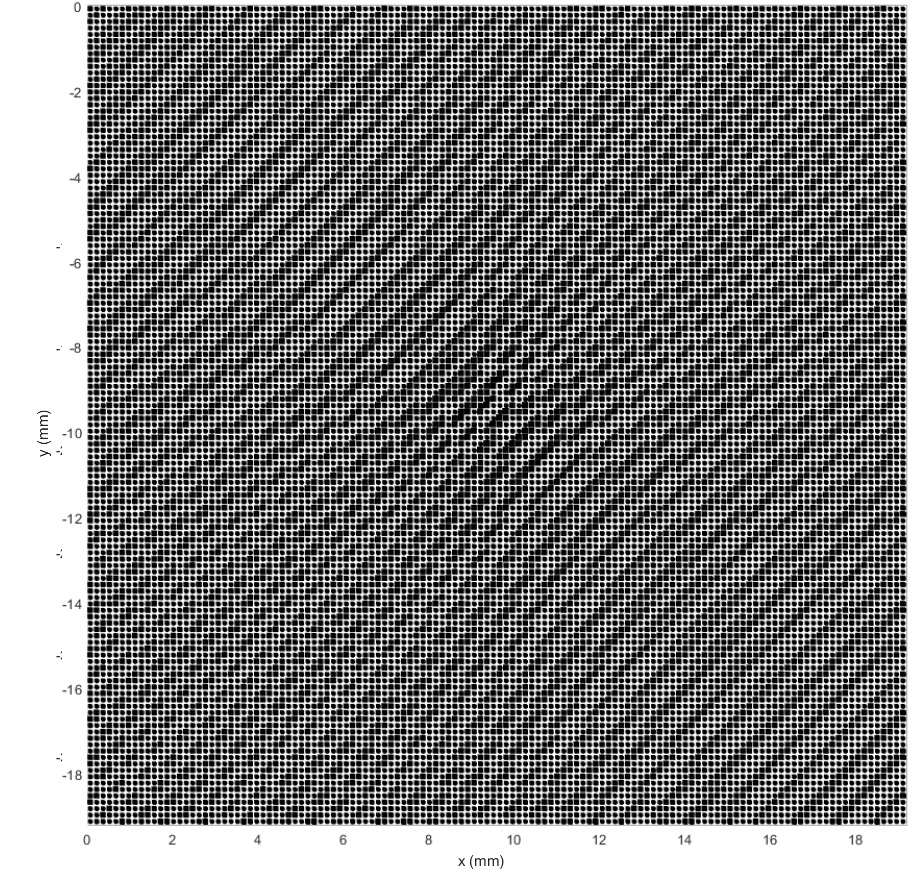}
\caption{Holographic metasurface of the CGH of Bessel beam implemented using unit cells with gaps variation.}
\label{f5}
\end{figure}

\begin{figure}[H]
\centering
\includegraphics[width=8cm]{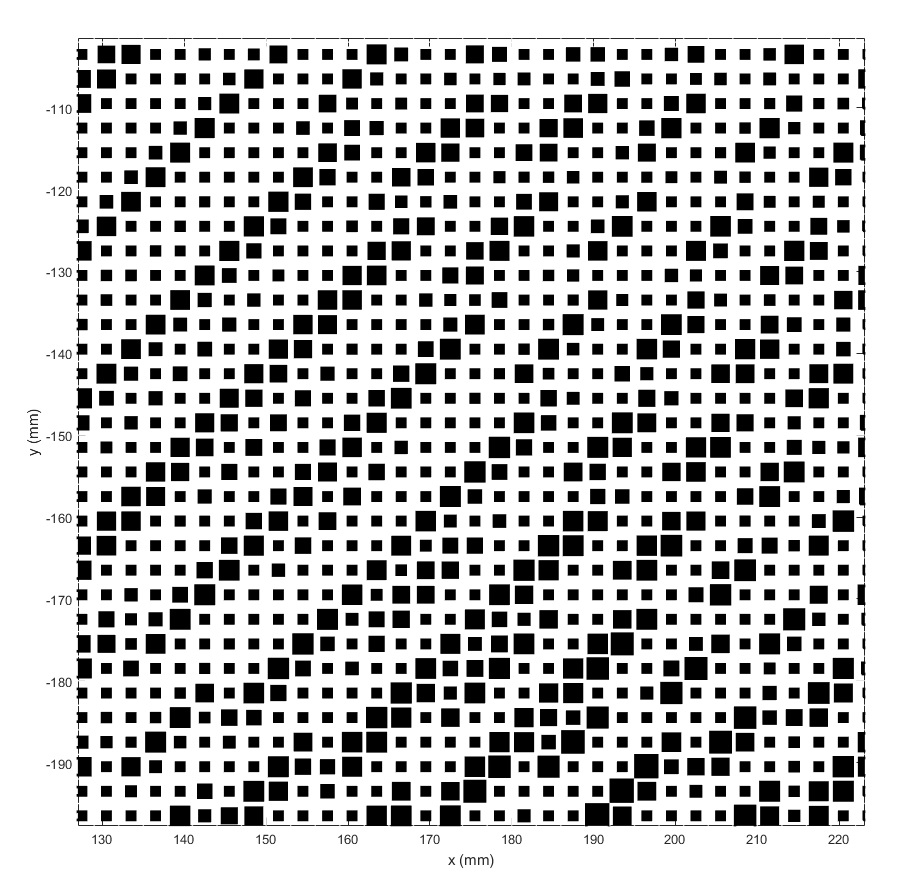}
\caption{Detail of the central region of the holographic metasurface of a non-diffracting beam by surface impedance.}
\label{f2a}
\end{figure}

\begin{figure}[H]
\centering
\includegraphics[width=8.1cm]{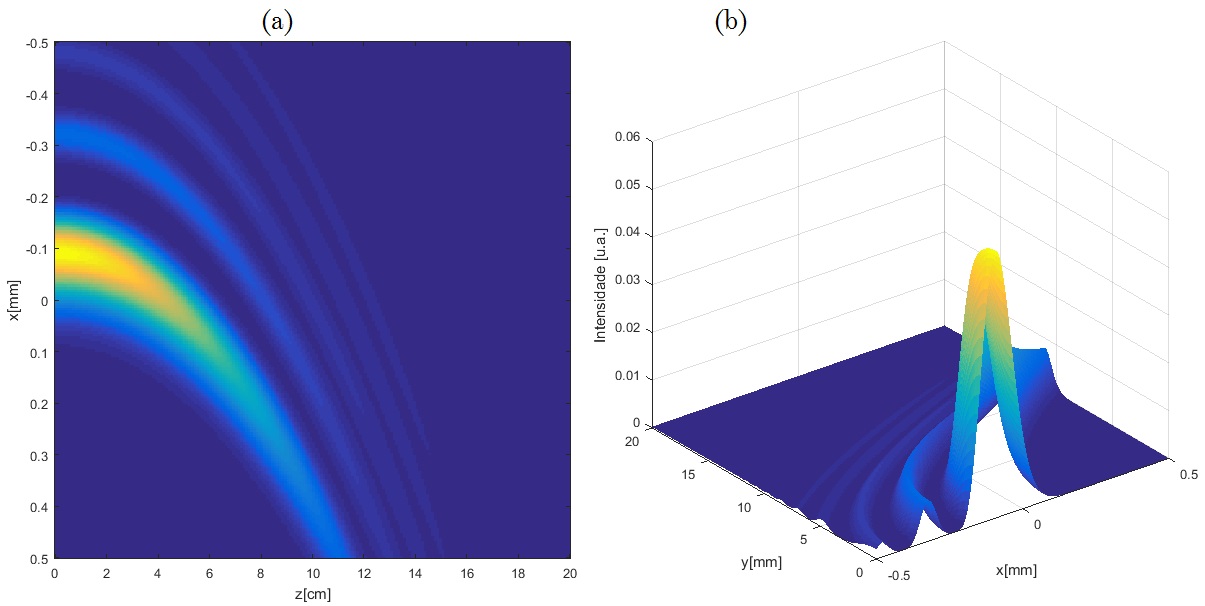}
\caption{(a) Orthogonal projection, and (b) transverse profile of the Airy beam.}
\label{f5b}
\end{figure}

\begin{figure}[H]
\centering
\includegraphics[width=5cm]{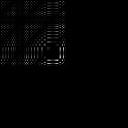}
\caption{Image of the computer-generated hologram of a Airy beam with resolution of 128x128 pixels and at wavelength $\lambda=12.33$ mm.}
\label{f6}
\end{figure}

\begin{figure}[H]
\centering
\includegraphics[width=9cm]{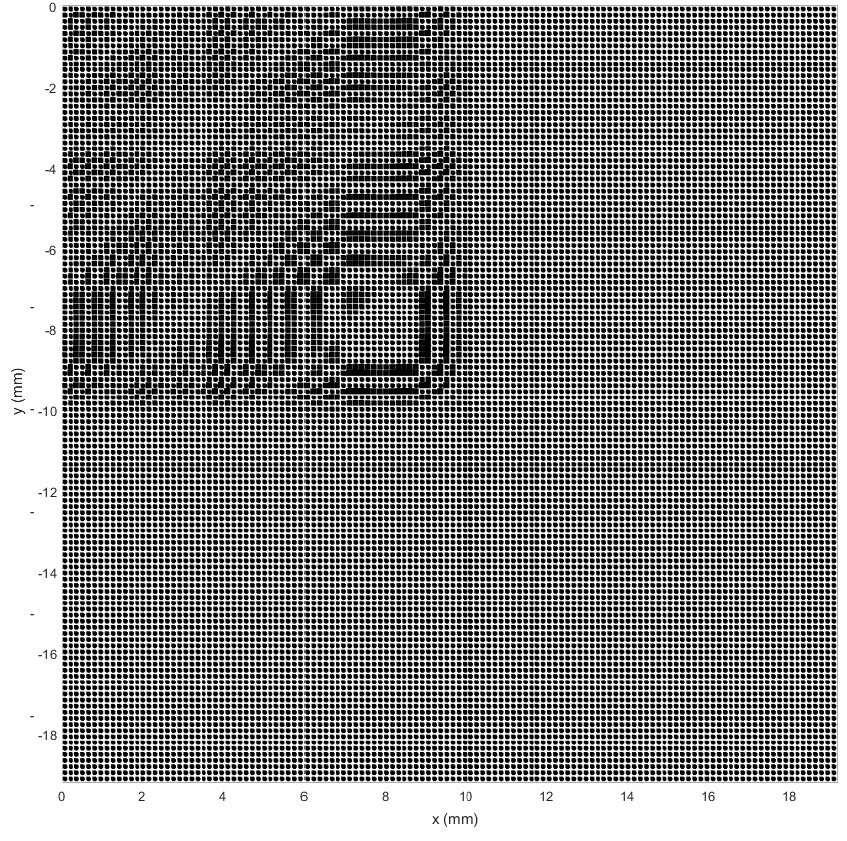}
\caption{Holographic metasurface of the CGH of Airy beam implemented using unit cells with gaps variation.}
\label{f7}
\end{figure}

\begin{figure}[H]
\centering
\includegraphics[width=7cm]{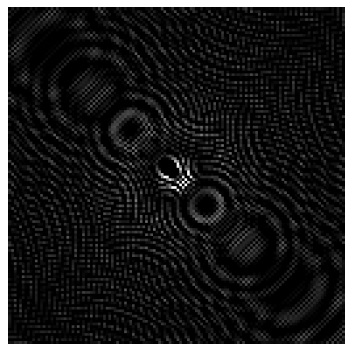}
\caption{Image of the computer-generated hologram of a FW beam with resolution of 128x128 pixels and at wavelength $\lambda=12.33$ mm.}
\label{f8}
\end{figure}

\begin{figure}[H]
\centering
\includegraphics[width=12cm]{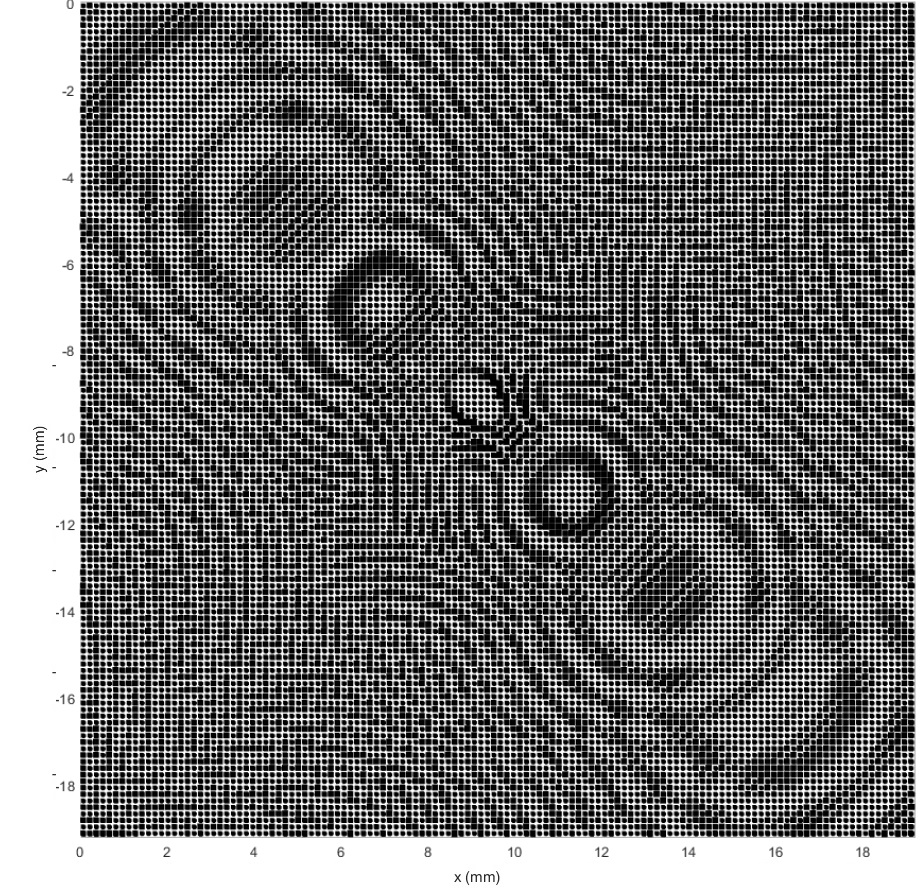}
\caption{Holographic metasurface of the CGH of FW beam implemented using unit cells with gaps variation.}
\label{f9}
\end{figure}

For this first holographic metasurface, we reproduce the CGH of a Bessel beam of zero order, with a resolution of 128x128 pixels, and generated to a wavelength of $\lambda=1.11 \mu m$, corresponding to our operating frequency of $270THz$ (see Figures \ref{f4b} and \ref{f4}), the corresponding holographic metasurface is shown in the Figures \ref{f5} and \ref{f2a}.  \\

we also reproduce the CGH of a Airy beam, We have an Airy beam with decay parameter a = 0.1, with a resolution of 128x128 pixels, and generated to a wavelength of $\lambda=1.11 \mu m$, corresponding to our operating frequency of 270THz (see Figures \ref{f5b} and \ref{f6}), the corresponding holographic metasurface is shown in the Fig. \ref{f7}. And, the CGH of a FW beam, with a resolution of 128x128 pixels, $N=6$, spot size of XX mm and generated to a wavelength of $\lambda=1.11 \mu m$, corresponding to our operating frequency (see Fig. \ref{f8}), the corresponding holographic metasurface is shown in the Fig. \ref{f9}. \\

\begin{figure}[H]
\centering
\includegraphics[width=9cm]{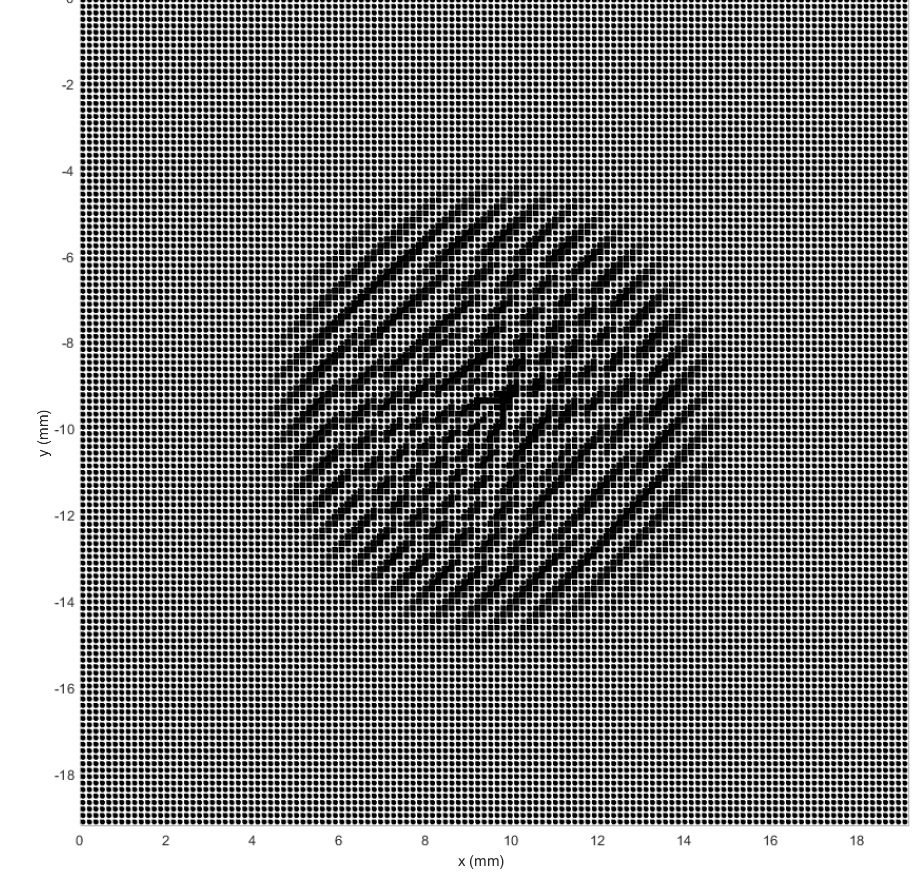}
\caption{Holographic metasurface of the CGH of LG beam implemented using unit cells with gaps variation.}
\label{f7b}
\end{figure}

\begin{figure}[H]
\centering
\includegraphics[width=9cm]{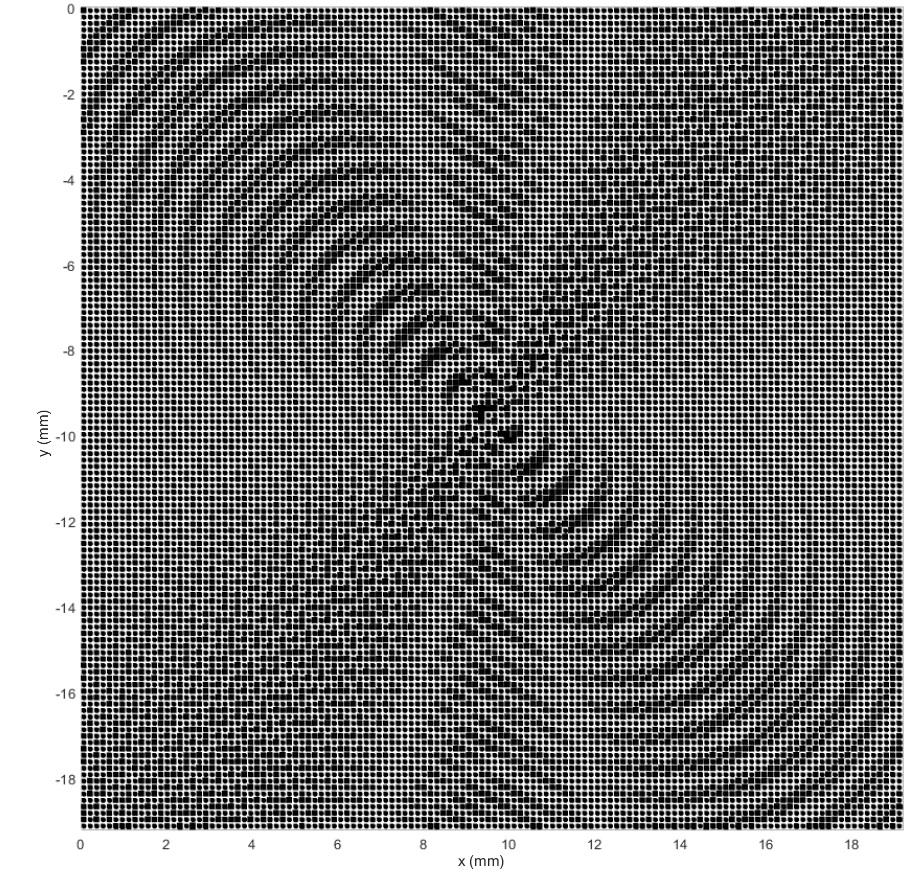}
\caption{Holographic metasurface of the CGH of Mathieus beam implemented using unit cells with gaps variation.}
\label{f8b}
\end{figure}

Others HMS for LG beams and Mathieus beams (Figures \ref{f7b} and \ref{f8b}, respectively) ~\cite{ref:36}.

\begin{figure}[H]
\centering
\includegraphics[width=16cm]{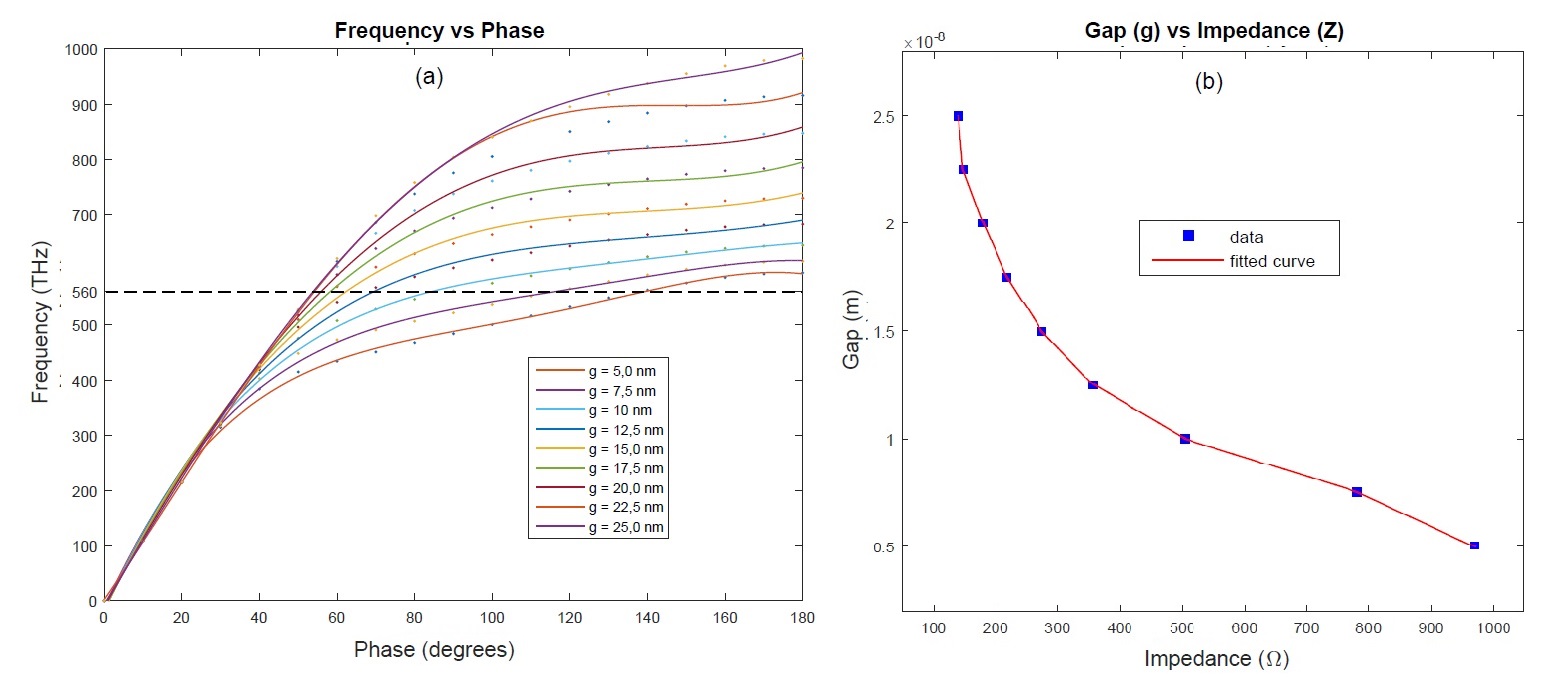}
\caption{(a) Variations of frequency with phase for each value of gap ($g$), the operating frequency was defined to 24.34 GHz. (b) Variation of values of gaps according to surface impedances and its corresponding adjustment curve.}
\label{f10}
\end{figure}

\h The second holographic metasurface works at operating frequency of 560THz ($\lambda = 536nm$), the metal used was Aluminum (Al) on a substrate of Silicon (Si) with permitivity $\epsilon=11.9$ and thickness $t=20.5$nm. The lattice parameter was set at $d=75$nm, and gaps were setted from $g=5$nm till $g=25$nm with an interval of 2.5nm. The frequency was defined from 0 till 1000THz. In the four sides of the unit cell (see Fig. \ref{f1}) were set periodic boundary conditions and PEC for the propagation direction. The graph of the dispersion curves for each value of gap and variation of gaps with surface impedance are shown in the Figures \ref{f10} and \ref{f11} respectively. After making the adjustment of the curve of $g$ vs $Z$, we obtained the interval for surface impedance: $[Z_{\textrm{min}}, Z_{\textrm{max}}] = [140.7\Omega, 970.1\Omega]$, and we also found the following modulate values: $X=140.7\ \Omega$, $M=132$.

\begin{figure}[H]
\centering
\includegraphics[width=9cm]{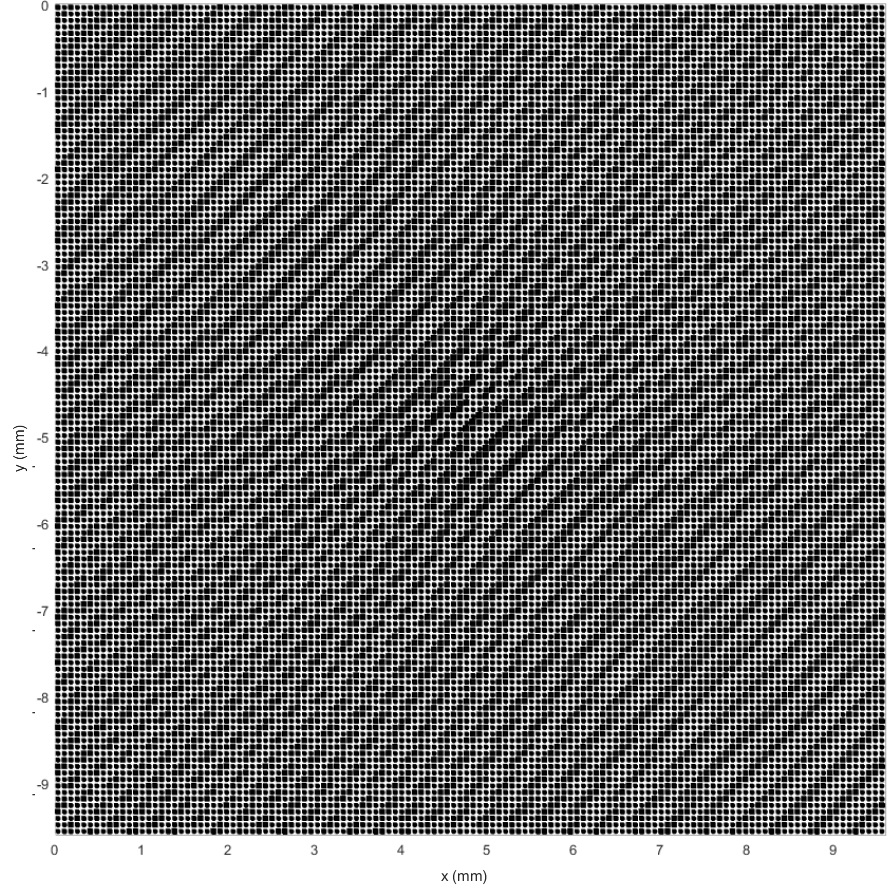}
\caption{Holographic metasurface of the CGH of Bessel beam implemented using unit cells with gaps variation.}
\label{f12}
\end{figure}

\begin{figure}[H]
\centering
\includegraphics[width=10cm]{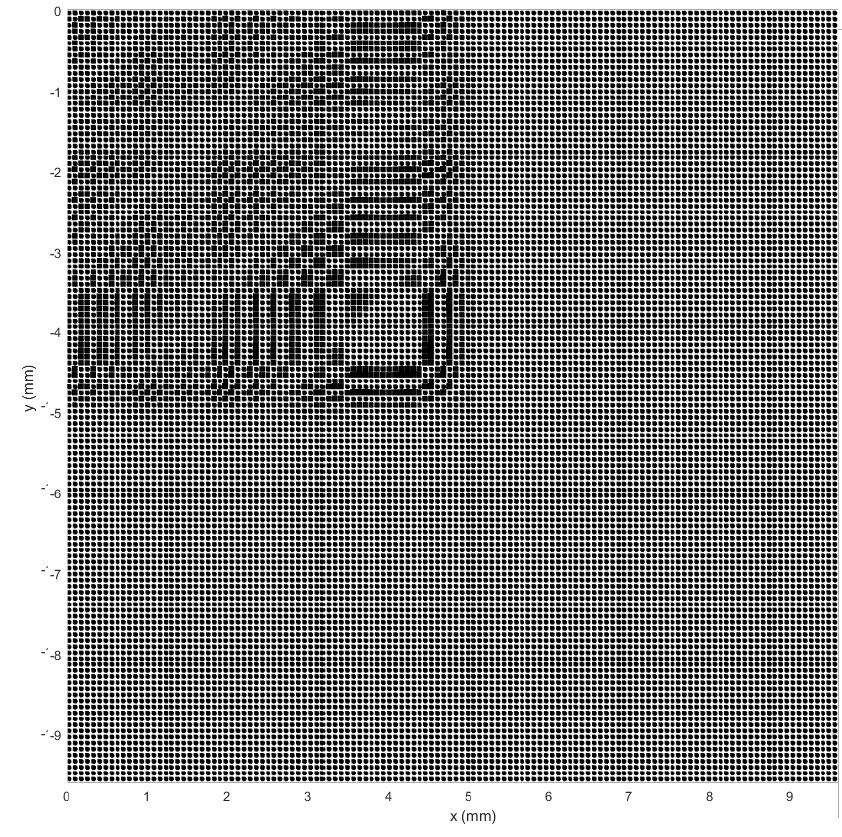}
\caption{Holographic metasurface of the CGH of Airy beam implemented using unit cells with gaps variation.}
\label{f13}
\end{figure}

\begin{figure}[H]
\centering
\includegraphics[width=10cm]{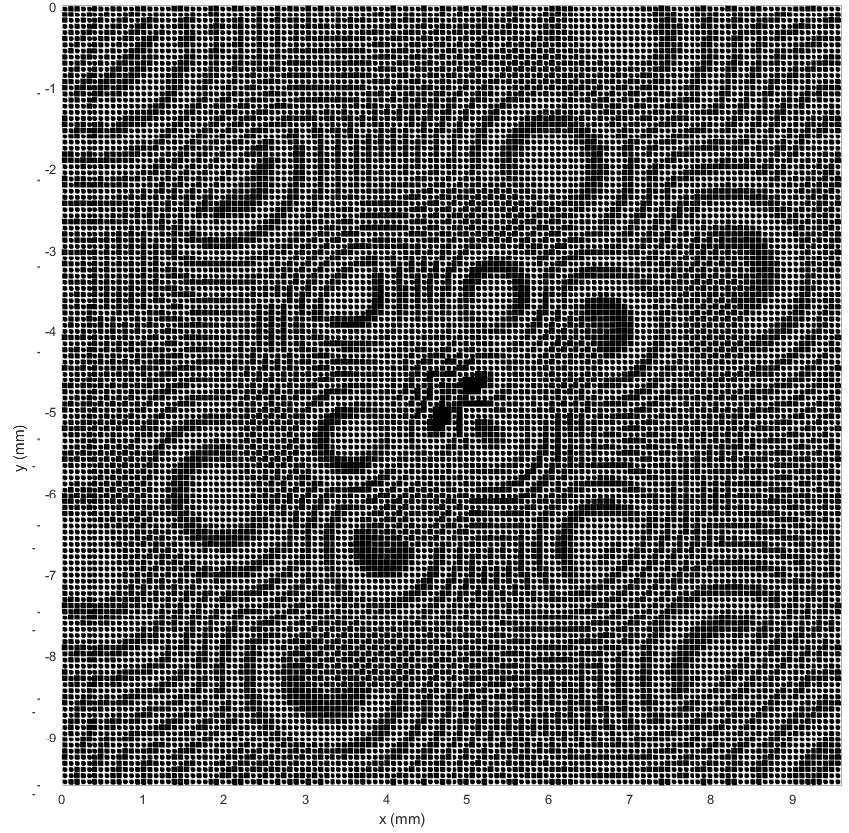}
\caption{Holographic metasurface of the CGH of FW beam implemented using unit cells with gaps variation.}
\label{f14}
\end{figure}

\begin{figure}[H]
\centering
\includegraphics[width=9.5cm]{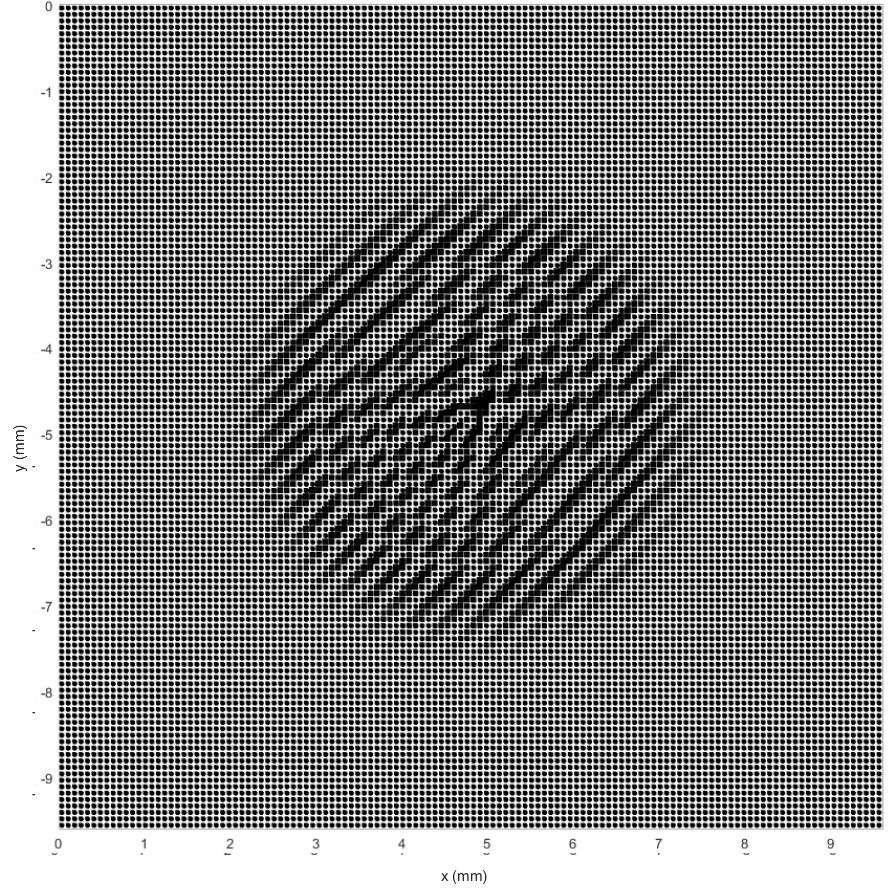}
\caption{Holographic metasurface of the CGH of LG beam implemented using unit cells with gaps variation.}
\label{f7c}
\end{figure}

\begin{figure}[H]
\centering
\includegraphics[width=9.5cm]{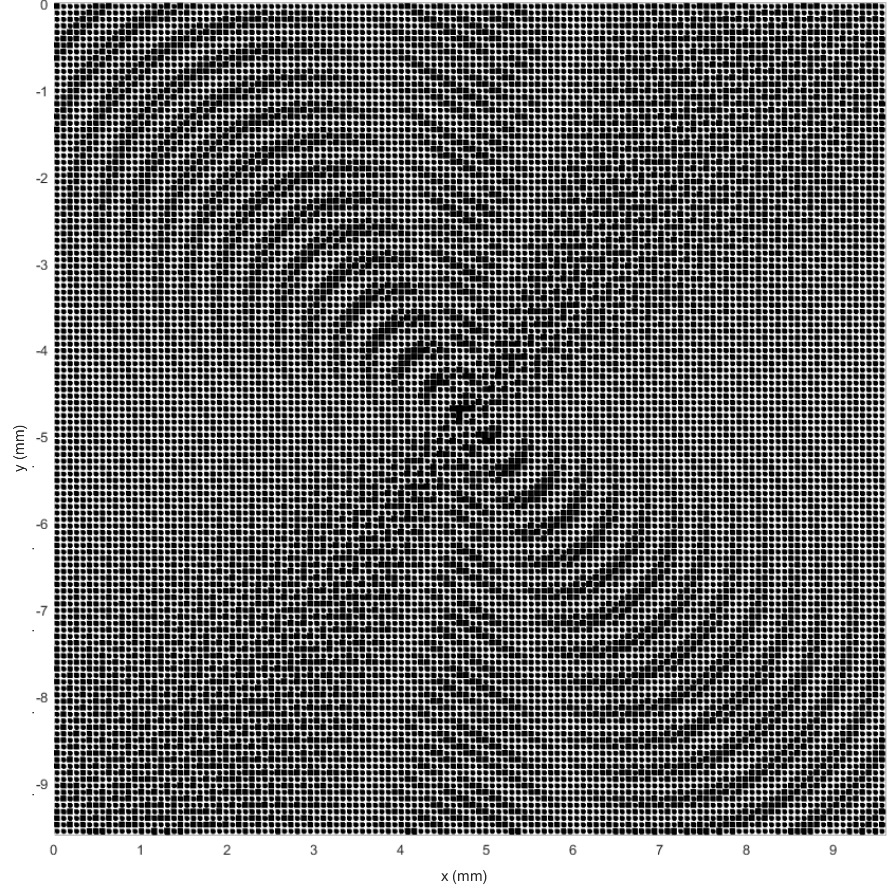}
\caption{Holographic metasurface of the CGH of Mathieus beam implemented using unit cells with gaps variation.}
\label{f8c}
\end{figure}

For this second holographic metasurface, we reproduce the CGH of a Bessel beam of zero order, with a resolution of 128x128 pixels, and generated to a wavelength of $\lambda = 536nm$, corresponding to our operating frequency of 560THz, the corresponding holographic metasurface is shown in the Fig. \ref{f12}. We also reproduce the CGH of a Airy beam, with a resolution of 128x128 pixels, and generated to a wavelength of $\lambda = 536nm$, corresponding to our operating frequency, the corresponding holographic metasurface is shown in the Fig. \ref{f13}. And, we also reproduce the CGH of a FW beam, with a resolution of 128x128 pixels, $N=6$, spot size of 47.9 mm and generated to a wavelength of $\lambda = 536nm$, corresponding to our operating frequency, the corresponding holographic metasurface is shown in the Fig. \ref{f14}.Others HMS for LG beams and Mathieus beams (Figures \ref{f7c} and \ref{f8c}, respectively) ~\cite{ref:36}. \\

\h {\em 6. Conclusions} ---
This work presents a way for controlling and manipulating the electromagnetic radiation through the computational realization of holographic metasurfaces to generation of the non-diffracting waves. Holographic metasurfaces (HMS) are simulated by modeling a periodic lattice of metallic patches on dielectric substrates with sub-wavelength dimensions, where each one of those unit cells alter the phase of the incoming wave. The surface impedance (Z) allows to control the phase of a wave through the metasurface in each unit cell. The sub-wavelength dimensions guarantees that the effective medium theory is fulfilled. The metasurfaces are designed by the computer-generated hologram (CGH) of non-diffracting waves are generated and reproduced using such HMS in the optical frequencies. This work offers a great advantage of using metasurfaces instead of conventional spatial light modulators (LC-SLM) for generating optical beams with a much better optical resolution due to small dimensions of the unit cells of metasurfaces which behave as pixels of the LC-SLM. The results is according to the predicted by non-diffracting beams theory. These results are important given the possibilities of applications in optical tweezers, optics communications, optical metrology, 3D imaging, and others in optics and photonics. \\

\h {\em Acknowledgements} ---
The authors would like to thank Rafael A. B. Suarez for fruitful technical discussions. This work was supported by Federal University of ABC (UFABC); Coordenaçao de Aperfeiçoamento de Pessoal de Nivel Superior-CAPES; Fundaçao de Amparo a Pesquisa do Estado de Sao Paulo-FAPESP (16/19131-6); Conselho Nacional de Desenvolvimento Cientifico e Tecnologico-CNPQ (302070/2017-6). \\

\end{document}